# 1. Introduction

For specific application, a polymer cannot be selected by only considering its bulk properties because its surface state usually plays a vital role. Low density polyethylene (LDPE), widely used in food packaging industry, presents no specific chemical functionality on its surface. It can therefore be modified to enhance its adhesion properties so as to subsequently deposit barrier layers dedicated to the food packaging. The surface activation could be accompanied by a slight etching process, which improves the surface roughness, hence increasing its hydrophilicity and furthermore the adhesion of the subsequent layer.[1]

Plasma treatment is known as a very effective way to enhance the hydrophilicity of a polymer surface by only modifying the outermost atomic layers without affecting its intrinsic bulk properties.[2,3] Etching, cleaning, crosslinking, and grafting are the main interactions that may occur on polymer surfaces.[4] Low pressure and atmospheric pressure plasmas have found widespread use for the pretreatment of polymers in industrial applications. Plasmas, such as dielectric barrier discharges (DBD) and radio frequency (RF) discharges, gain much interest as processing techniques for the activation or the modification of polymer surfaces, even if they present some inconvenients.[5–7] Chen et al.[8] described not only the activation but also the deterioration of the polytetrafluoroethylene (PTFE) surface after its exposure to an RF plasma powered at 100W during 100 s. However, no or minor damage was noticed when the films were treated in the remote afterglow in the same plasma conditions. The damage of poly(vinyl chloride) surface treated in an RF discharge[9] or in DBD discharge[10] was evidenced.

Many studies[11–13] underlined that a DBD treatment induced a surface oxidation of polyethylene (PE). Gilliam et al.[14] reported the treatment of LDPE in a low-temperature cascade arc plasma for a flow rate of 500–3000 sccm with an RF power set at 8W and a pressure of 50 mTorr. They underlined that a water contact angle (WCA) of 40° was reached (94° before treatment) after 20 s of argon plasma treatment while adding a reactive gas as oxygen resulted in greater damage of the LDPE surface. These authors demonstrated that the film was modified not only on the topmost surface layers but also in the bulk.

Plasma discharges, rich in active species (ions, electrons, and radicals), usually lead to a strong damage of the polymer surface. On the contrary, the post-discharge is much less aggressive in that most of the electrons and ions are neutralized upstream. As the radicals present a longer lifetime than those of the charged species, they can travel through the post-discharge and reach the polymer surface.

Plasma surface modification of polymers is well known with an emphasis on the replacement of C–H bonds by polar, oxygen containing new groups. However, much less has been reported so far on polymer bulk modification induced by plasma and in particular on the oxygen diffusion toward the bulk. The modified depth has been determined to be typically comprised between 50 and 500 Å.[15] In the case of high density, PE and LDPE films modified in an $Ar^+$ plasma discharge for exposure times varied from 0 to 400 s, with a power of 1.7W, Svorcik et al.[16] showed by RBS method that the diffusion depth of the incorporated oxygen lay 25 nm beneath the sample surface in both polymer types after 250 s of treatment.

Our article is focused on the activation of LDPE surfaces induced by the flowing post-discharge of an atmospheric RF plasma torch. The post-discharge is supposed to modify the surface and bulk of the polymer without damaging the topmost surface layer. The novelty of this work lies on the characterization of the oxygen diffusion depth by using angle-resolved (AR)-XPS and time-of-flight (ToF-SIMS) as complementary techniques. Other results obtained from WCA measurements and Fourier transform infrared (FTIR) spectroscopy consolidate the conclusions provided by AR-XPS and ToF-SIMS.

## 2. Experimental details

### 2.1. Materials

The LDPE polymer films used for this study were provided by PackOplast. They were 37% crystalline and presented a thickness of 40 µm. For each series of experiments, three LDPE films were used so as to evaluate the uncertainties.

### 2.2. Configuration: plasma torch and parameter conditions

The polymer was treated using an AtomfloTM-250D plasma source from SurfX Technologies LLC.[7,17] The source was connected to a gas supply and an RF power generator with a matching network operating at 27.12 MHz. According to the specifications of the manufacturer, the power could be tuned between 60 and 90 W, the carrier gas could only be argon for flow rates ranging between 20 and 40 $L.min^{-1}$, whereas the reactive gas could only be oxygen for flow rate comprised between 0 and 25 $mL.min^{-1}$. In our case, the argon flow rate was fixed at 30 $L.min^{-1}$ for all the experiments. The treatment was achieved with a source-to-substrate distance (gap) of 9 mm.

### 2.3. Water contact angle measurements

The wettability modifications of the polymer surfaces were followed by dynamic WCA measurements by means of a Krüss DSA 100 (Drop Shape Analysis system, Krüss Gmbh, Germany). These measurements were performed in an air-conditioned room, and the working liquid was milli-Q water. Advancing and receding angles were measured by depositing a drop of 5 μL on the surface, then increasing its volume to 15 μL, finally decreasing it. The reported advancing WCA was the maximum angles observed during the drop growth. The values of the receding WCA were those measured just before the contact surface reduction, i.e., the last measurements (from 5 to 7) before the distortion of the drop. Each advancing WCA presented in this article is an average of over five drops deposited, with a maximum error estimated to 3°.

## 2.4. SEM

Scanning electron microscopy was performed using a 200 CX Jeol microscope at 200 kV. As the samples were insulating materials, a thin layer (~10 nm) of gold was deposited onto their surface in order to compensate the charge effect.

## 2.5. XPS

X-ray photoelectron spectroscopy measurements were performed with a PHI 5600 photoelectron spectrometer system with a Mg Kα X-ray source (1253.6 eV) operating at 300W. For each spectrum, 10 accumulations were acquired, and the pass energies were 93.90 eV for the wide spectra and 23.5 eV for high resolution spectra. All spectra were referenced to the C 1s peak of the carbon atom at 285.0 eV. The detection angle with respect to the surface was 45°. The surface analysis was carried out directly after the plasma treatment. High resolution XPS analyses were performed on the C 1s peak. The elemental composition was calculated after removal of a Shirley background and using the sensitivity coefficients provided by the manufacturer's handbook: $S_C$=0.205 and $S_O$=0.63.[18] The FWHM was set to 1.5 eV for all spectra components. For the AR-XPS measurements, the detection angle with respect to the surface was varied for the following several values: 18°, 25°, 35°, 45°, 55°, 65°, 75°, 85°, and 90°. The angular acceptance and the number of angular channels were 60° and 120°, respectively. The distance between two lines was 5°. The data were acquired with the AugerScan software (RBD Instruments, Incorporation, USA) and the peak fittings were achieved with the Casa XPS software (Casa Software Ltd, USA).

## 2.6. Fourier transform infrared in attenuated total reflection (ATR) spectroscopy

The composition changes of the polymer surfaces were analyzed by ATR infrared spectroscopy. The FTIR spectra were determined by means of a Nicolet 5700 FTIR spectrometer (Thermo Electron Corporation, MA, USA) equipped with an MCT (mercury cadmium telluride) detector cooled by liquid nitrogen, using a diamond/zinc selenide crystal. To ensure reproducible contact between the crystal faces and the polymer, the same pressure was applied to the crystal holder by means of a calibrated torque screw driver. Sixty-four scans with a resolution of 4 cm$^{-1}$ were carried out for the untreated and each one of the plasma-treated samples.

The infrared beam, in ATR spectroscopy, passes through the ATR crystal and penetrates the sample as an evanescent wave. According to [19,20], the penetration depth of the ketone CO stretching vibration (between 1720 and 1740cm$^{-1}$) can be estimated at 1180 nm. The peak fitting was realized with the PeakFit software (Systat Software, Inc., IL, USA).

## 2.7. ToF-SIMS

Depth profiling experiments were accomplished with a ToF-SIMS IV instrument from ION-TOF GmbH (Münster, Germany) equipped with two ion guns for analysis and etching. The analysis was performed using a pulsed 15 keV $Ga^+$ ion gun, whereas the etching was achieved using $Cs^+$ ions at the energy of 750 eV.

The analysis was rastered over a 61.5 × 61.5 µm$^2$ area centered inside the etching crater rastered over a 300 × 300 µm$^2$ area. The incident angle for both beams was 45°. A non-interlaced mode was used in order to limit the $Ga^+$ influence and consequently, to reduce significantly the ion-induced damage. SIMS depth profiles were performed exclusively in negative ion polarity. An electron flood gun was used for charge compensation during all depth profiles. The spectrum normalization, achieved for all the results presented in this work, refers to the normalization of each spectrum in the depth profile to its own total secondary ion intensity in the m/z = 0–60 range for the negative ion mode. This normalization method was used to facilitate the measurement of sputtering rates from the depth profiles.

Because of its too large thickness (40 µm), sputtering an entire LDPE film would require too much time for evaluating the etching depth related to the etching time. In addition, for specific experimental conditions ($Cs^+$ energy, current, and raster size), the heterogeneous crater resulting from the erosion makes the depth measurement almost impossible by profilometry technique. As an alternative, a thinner LDPE film was achieved by using polymer granules, from INEOS-Belgium, with a density of 0,933 g.cm$^{-3}$ and a molecular weight of 240 000 g.mol$^{-1}$. They were ground to obtain a powder of LDPE, which was dissolved in hot (about 150 °C) decalin (99%, Sigma–Aldrich) for 3 h. Then, the solution was deposited on a clean Si(100) for the spin coating process. By varying polymer concentration, spin speed and temperature, the thickness of the polymer films could be controlled to obtain thin LDPE film. The LDPE layer thickness was estimated simply by scratching the layer and measuring the step height with a profilometer. An average thickness of 120nm was measured by repeating this procedure several times. This film was then sputtered and analyzed by ToF-SIMS. The sputtering yield was calculated on the basis of the fluence of the $Cs^+$ ions (3.0 × 10$^{17}$ ions cm$^{-2}$) and the sputtering time. The etching rate was estimated to 82 Å.min$^{-1}$, hence permitting the estimation of the oxygen diffusion into the treated LDPE.

## 2.8. Optical emission spectroscopy (OES)

Optical emission spectroscopy was performed with a SpectraPro® 2500i from Acton research, Princeton instruments, Acton MA, USA equipped with a CCD camera (Princeton digital camera with 400 × 1340 pixels). For any OES measurement presented in this article, an LDPE sample was always placed under the plasma torch to a distance of 9mm, and the optical fiber was placed perpendicularly to the plasma torch and the sample. The data were collected with the software WinSpect32 (Princeton instruments, Acton MA, USA). A total of 1800 grooves/mm grating blazed at 500 nm (linear dispersion of 0.846mm at a wavelength of 579 nm) was employed to acquire spectra in the visible/near infrared range. The exposure time was fixed to 25ms, and 50 accumulations were achieved for each spectrum acquired.

# 3. Results

## 3.1. Discharge treatment versus post-discharge treatment

In order to evidence the advantage of using a post-discharge rather than other plasma sources, the treatment of LDPE was also achieved in a DBD under similar conditions (60W of power, 30 s of treatment, atmospheric pressure, circular electrodes with 40mm of diameter and alumina dielectric with 2mm of thickness). Figure 1 shows (a) SEM micrographs of untreated LDPE, (b) LDPE treated by post-discharge, and (c) LDPE treated by DBD. The first image (a) is dominated by irregular protrusions on the surface, which can also be observed on the image of the film treated by the plasma torch (b), thus evidencing that the plasma treatment did not induce a significant modification or strong damage of the LDPE surface. On the contrary, the topography of the image (c) shows the appearance of microscale protrusions due to a strong surface etching, attributed to the DBD. A similar behavior was observed when the PE film surface was modified by DBD in the air.[11]

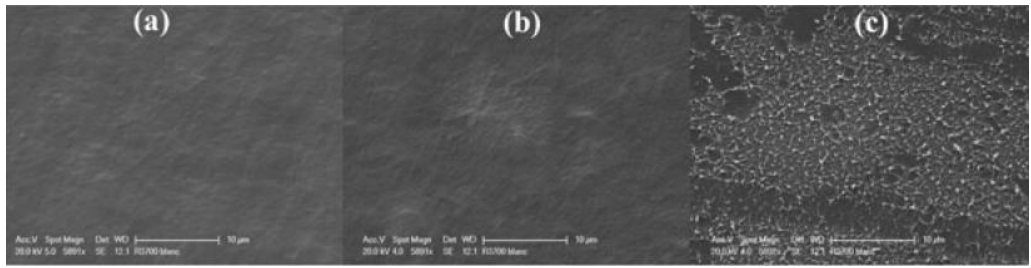

Figure 1. SEM micrographs of (a) untreated LDPE; (b) LDPE treated 60 s in the post-discharge; and (c) LDPE treated 30 s by DBD.

## 3.2. Wettability and morphology of the surface

As expected from the state of art [12], the plasma torch treatment causes a decrease in the hydrophobic character of the LDPE films, the films treated by Ar–$O_2$ plasma being even more hydrophilic. Figure 2a illustrates the variation of the advancing WCAs of the LDPE films versus the treatment time in the range of 0–300 s for several plasma powers. It can be observed that the contact angle of the untreated LPDE sample (94.2°) decreases after an Ar or Ar-$O_2$ post-discharge treatment. For an exposure of 30 s to the Ar post-discharge powered at 60 W, the contact angle decreases from 94° to 56° (Fig. 2a) instead of 47° for a sample treated under plasma containing oxygen (25 mL.min$^{-1}$) (Fig. 2b). For a treatment time of 30 s, the results of WCA confirm that a saturation of the hydrophilic groups is reached on the LDPE surface. Comparable behaviors were observed for LDPE treated with other plasma configurations and plasma gases.[21–24] Similar contact angles (i.e., 45°–60°) were reported for high density polyethylene (HDPE) treated with a DBD operating in air.[12] Figure 2b shows also a progressive increase in the hydrophilic character of the surface for treatment times higher than 30 s. The presence of oxygen mixed with the carrier gas allows a decrease in the WCA of at least 10°, whatever the value of the $O_2$ flow rate (5, 15, or 25 mL.min$^{-1}$).

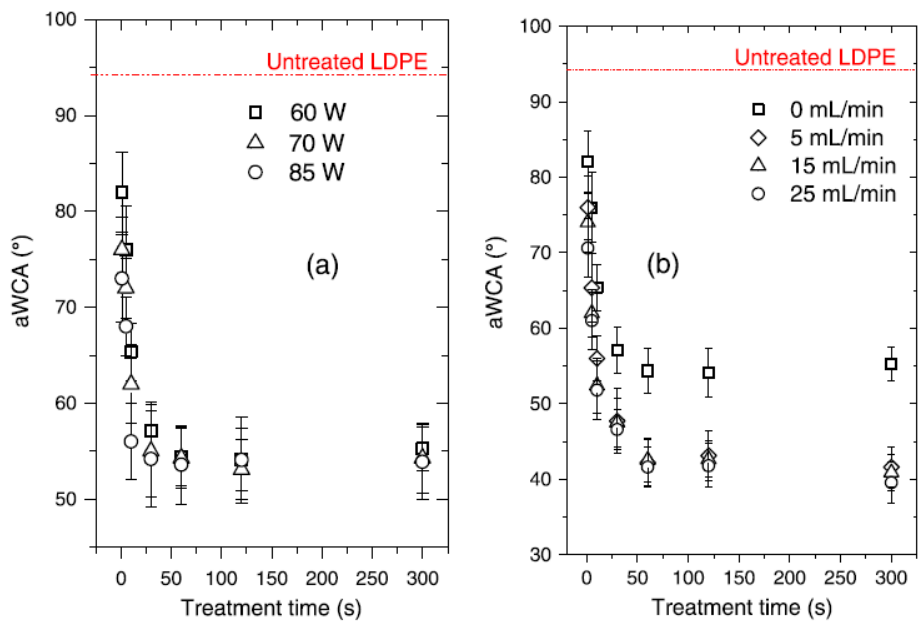

*Figure 2. Advancing water contact angle (AWCA) of treated LDPE films in terms of (a) the treatment time with pure argon plasma at different powers and (b) the treatment time with different oxygen flow rates at 60W of plasma power.*

### 3.3. Modification of the outermost LDPE surface

The modification of the surface chemistry was investigated by XPS. The wide XPS spectrum of native LDPE surface shows an O 1s peak (4.7%) revealing a slight oxidation of the surface due to the manufacturing process of LDPE. An increase in the oxygen surface concentration is detected on the spectrum of the treated film.

Figure 3a,b,d illustrates the chemical composition ratio O/C of the top surface of LDPE with different treatment times, plasma powers, and oxygen flow rates. The increase of the O/C ratio indicates that more oxygen is grafted on the surface of the film with the increasing time of treatment (Fig. 3a). Figure 3b shows that the O/C ratio increases slowly with the increasing power. This increase is not in linear dependence with the cosine of static WCA, as shown in Fig. 3c. Vandencasteele et al.[25] showed the linear dependence between the cosine of WCA and the percentage of fluorine detected on the PTFE surface. The nonlinearity, observed in our case, could be explained by an oxygen diffusion toward the bulk, because the analysis depth of the XPS is around 5–10nm or by a smooth etching process that occurs on the surface of the polymer. Figure 3d shows the increase in the O/C ratio with as a function of the oxygen flow rate. This increase may come from an increase in the oxygen concentration onto the surface and in particular in the bulk close to the topmost surface layers.

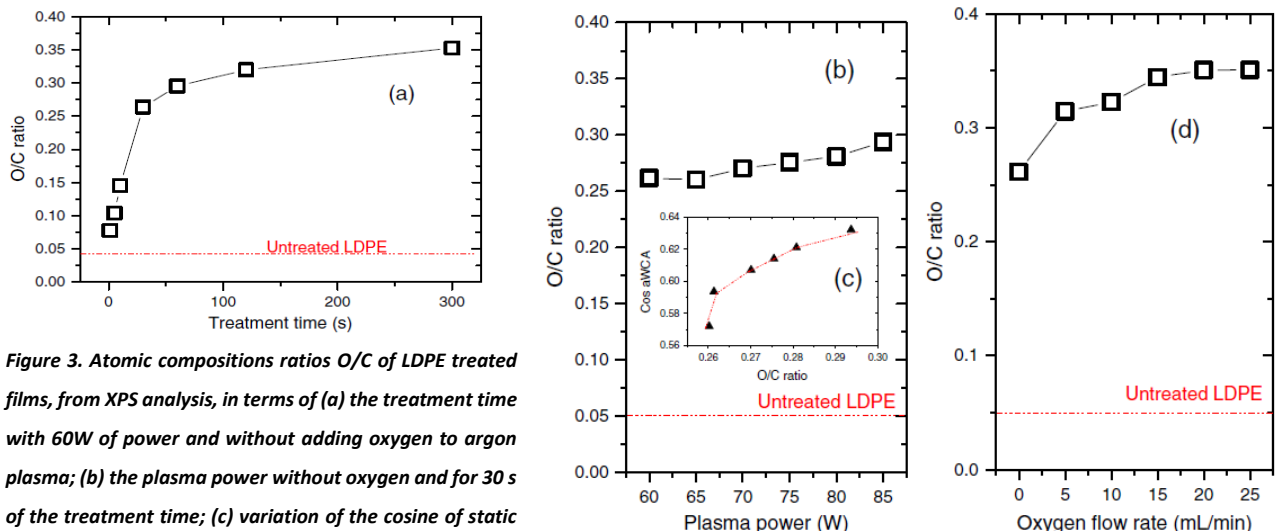

*Figure 3. Atomic compositions ratios O/C of LDPE treated films, from XPS analysis, in terms of (a) the treatment time with 60W of power and without adding oxygen to argon plasma; (b) the plasma power without oxygen and for 30 s of the treatment time; (c) variation of the cosine of static WCA in terms of O/C ratio for different plasma powers; and (d) the oxygen, mixed to argon, flow rate with 60W of power and during 30 s.*

### 3.4. Chemical and structural modifications of the bulk

Plasma surface modifications are confined only to a few tens of nanometers below the surface. FTIR spectroscopy in ATR mode is among the methods that can be used to analyze the bulk, i.e., few hundreds of nanometers.[26] The spectra are normalized with the hydrocarbon stretching vibration line at 2916 cm$^{-1}$ to compare the intensities of these lines with the spectra of different samples. Band assignments (taken from literature) are listed in Table 1.

| Peak no. | Wavenumber (cm$^{-1}$) | Mode assignment |
|---|---|---|
| 1 | 1090–1150 | C–O stretch (C–OH) |
| 2 | 1300–1360 | O–H in plane deformation or COO$^-$ |
| 3 | 1463–1473 | C–H deformation (–CH$_2$–) |
| 4 | 1640 | C O stretch of hydrogen, bonded carboxylic acid |
| 5 | 1730–1740 | C O stretch of ketone |
| 6 | 2848 | C–H symmetric stretch (–CH$_2$–) |
| 7 | 2916 | C–H asymmetric stretch (–CH$_2$–) |
| 8 | 3200–3600 | O–H stretch (C–OH) |

*Table 1. Fourier transform infrared spectral mode assignment for low density polyethylene[11,25,26]*

Figure 4 shows the FTIR spectra of the untreated and plasma-treated LDPE. The absorption peaks at 2916 (no. 7), 2848 (no. 6), 1472 and 1464 (no. 3) cm$^{-1}$ are attributed to the methylene asymmetry stretch vibration, methylene symmetry stretch vibration, methylene symmetry rocking vibration, and methylene asymmetry changing angle vibration, respectively (cf. Table 1).

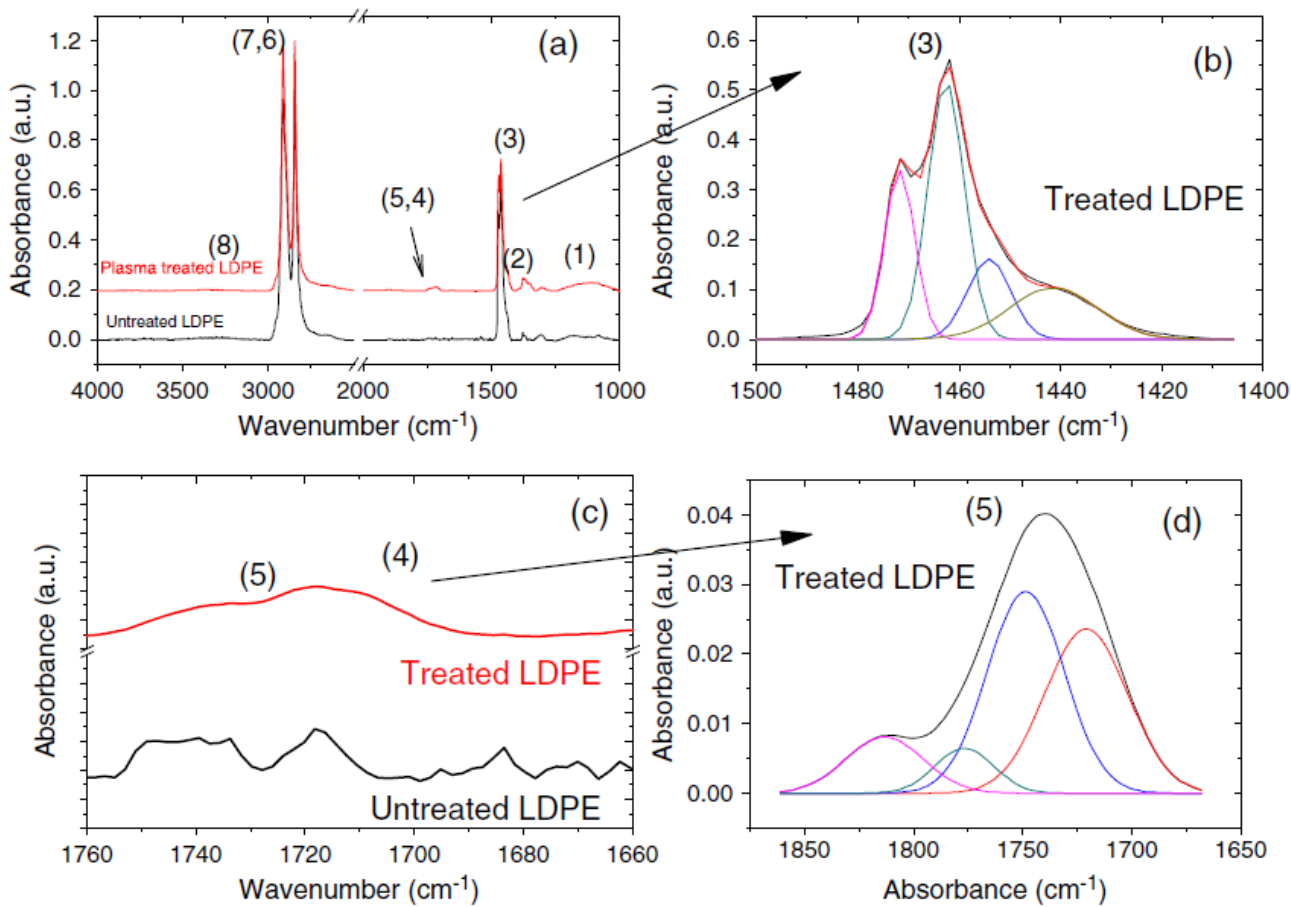

Figure 4. Infrared spectra of (a) untreated and treated LDPE (30 s at 60W without adding $O_2$) around 4000-1000 $cm^{-1}$; (b) the peak fitting of peak no. 3 between 1500 and 1400 $cm^{-1}$; (c) around 1780-1600 $cm^{-1}$; and (d) the peak fitting of peaks 4 and 5 (d).

Because of treatment, an additional peak (Figs. 4c,d) appears at 1737 $cm^{-1}$, which corresponds to the C O (no. 5) stretching vibration, and the band at 1640 $cm^{-1}$ corresponds to the COO (no. 4) asymmetrical stretching. For all samples, a weak shoulder is observed in the range 1350–1300 $cm^{-1}$, which can be attributed to $COO^-$ (no. 2). The v(O–H) vibrations are observed at 3400 $cm^{-1}$ (no. 8) related to hydroxyl, carboxyl, and peroxide groups, and the v(C–O) lines are observed at 1090 $cm^{-1}$ (no .1), which can be interpreted as the appearance of hydroxyl, carboxyl, ether, ester, and peroxide groups.

The peak fitting of the FTIR spectrum between 1500 and 1400 $cm^{-1}$ (Fig. 4b) shows two additional peaks at 1454 and 1440 $cm^{-1}$, corresponding to carbonyl groups (aldehyde or ketone). The C–OH and C O functional groups are generated on the LDPE surface, which was also confirmed by the XPS results.

The effects of the treatment time, plasma power, and oxygen flow rate on the surface/bulk polymer oxidation are compared by evaluating the integrated area of the peak at 1737 $cm^{-1}$. This peak can reveal an in-depth oxidation profile because the penetration depth of the ketone vibration C O at 1737 $cm^{-1}$ is 1180 nm. The variations of the integrated area of this C O peak are therefore depicted in Fig. 5 in terms of the plasma parameters (treatment time, power, and oxygen flow rate).

The integrated area of the C O vibration increases with the treatment time and reaches a plateau after 60 s of treatment (Fig. 5a). However, this peak increases linearly with the power or the oxygen flow rate, as introduced in Fig. 5b,c, thus suggesting a significant oxygen diffusion into the bulk. The C O vibration intensity variations complement the results of the O/C ratio calculated from the XPS analysis, because the analysis depth of the ATR-FTIR can reach ~1 µm, whereas it is only ~5–10nm in the case of XPS. The oxygen diffusion through the bulk is discussed in the next section in order to estimate as accurately as possible its penetration depth and the underlying mechanisms.

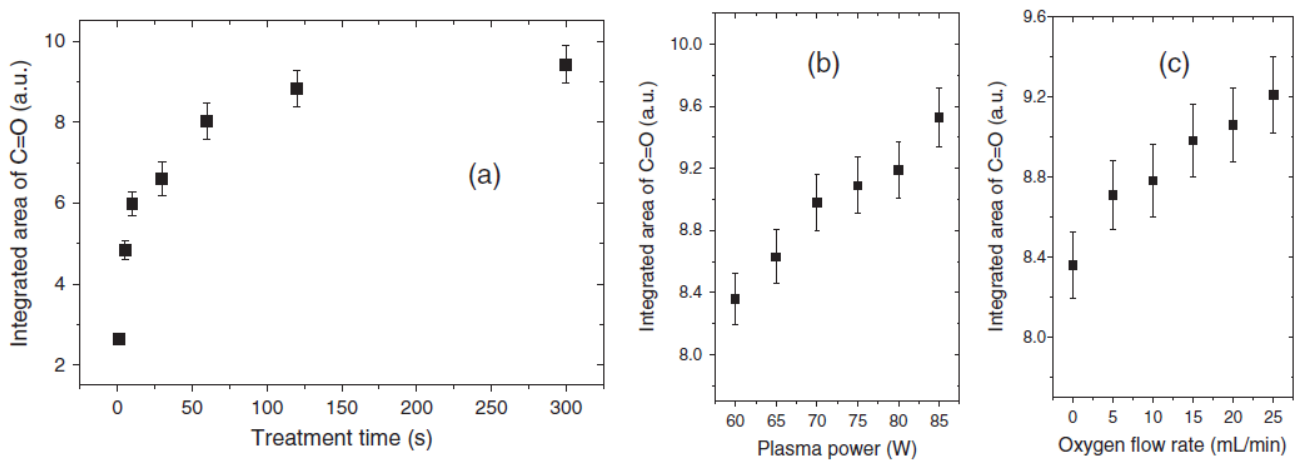

Figure 5. Variation of the integrated area of the peak C O at 1737 cm$^{-1}$, taken from FTIR analysis, as a function of (a) the treatment time; (b) the plasma power; and (c) the oxygen flow rate. The conditions of treatment are the same of those on Fig. 3a,b,d.

## 3.5. Study of the oxygen diffusion by ToF-SIMS

The diffusion of oxygen into the LDPE surface during the post-discharge treatment was studied by ToF-SIMS depth profiling. Polyatomic projectiles such as $C_{60}^+$ are widely used to perform depth profiling of organic layers. However, the sputtering of a polymer surface by $C_{60}^+$ ions is challenging as it can be crosslinked during the analysis. In this case, the ion signal recorded during the erosion drops is mostly because of cross-linking reactions, involving the formation of free radicals under the ion bombardment with a deposition of carbon into the matrix of the polymer.[27]

For these reasons, ToF-SIMS depth profiling was achieved with $Cs^+$ ions rather than $C_{60}^+$ or even $Ar^+$ ions. The $Cs^+$ ions are commonly used to achieve the elemental depth profiling of metal and semiconductor layers. Although less common, the feasibility of $Cs^+$ ions for depth profiling polymers has been recently demonstrated.[28,29] It is thought that the chemical reactivity of caesium plays a dominant role. The caesium implantation during depth profiling strongly enhances the negative ionization of the sputtered ions. Moreover, the implanted caesium prevents the cross-linking reactions in the polymer. Figure 6 shows a typical negative ion spectrum of an LDPE film (i) treated by an Ar–O$_2$ post-discharge powered at 60W during 30 s with $\Phi_{O2}$ = 25mL.min$^{-1}$ and (ii) sputtered by $Cs^+$ ions on a depth of 10 nm. The m/z peaks at 13, 17, 27, 41, 43, and 45 relative to the CH$^-$, OH$^-$, C$_2$H$_3^-$, C$_2$HO$^-$, C$_2$H$_3$O$^-$, and CHO$_2^-$ ions are characteristics of the modifications between untreated and treated LDPE samples.

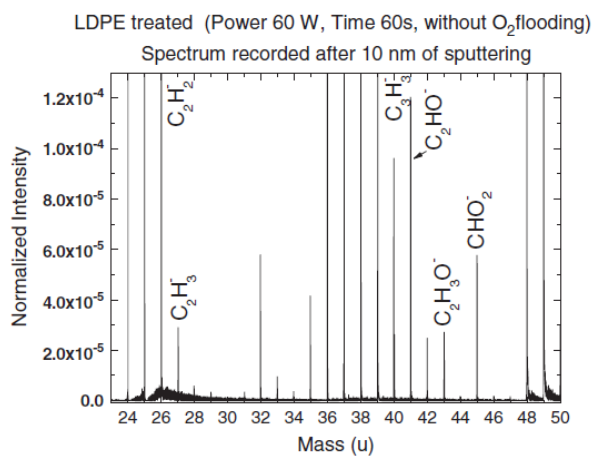

Figure 6. ToF-SIMS negative ion spectrum of treated LDPE (power 60 W, time 30 s, and $\Phi_{O2}$ 25 mL.min$^{-1}$) recorded after 10nm of Cs$^+$ sputtering.

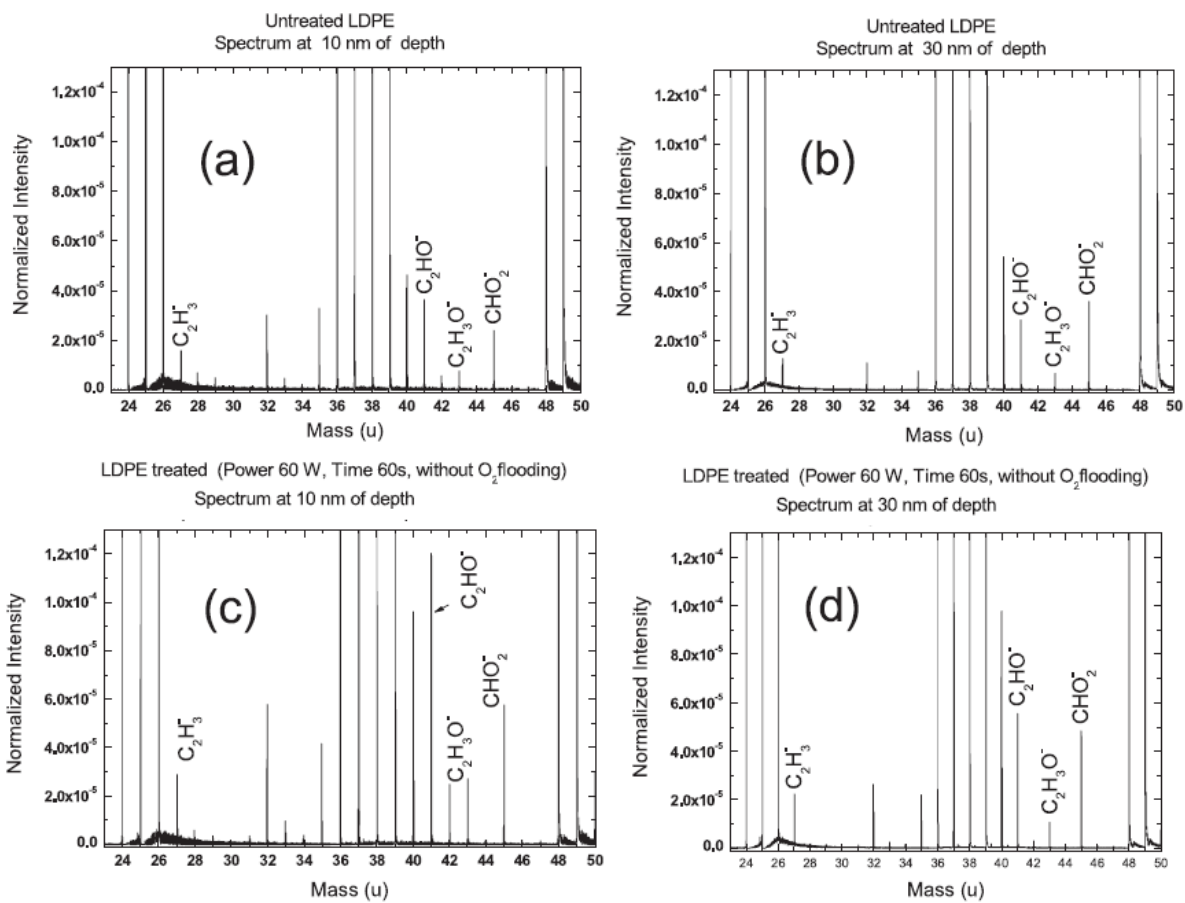

Figure 7. ToF-SIMS negative ion spectra of untreated and treated LDPE (power 60 W, time 30 s, and without adding oxygen) recorded at 10nm (a,c) and 30nm of depth (c,d).

The ToF-SIMS spectra presented in Fig. 7 were recorded for an untreated film and a film treated by a pure Ar post-discharge, powered at 60W for 30 s. In each case, the surface was probed 10 and 30 nm in-depth from the surface, after the sputtering induced by the $Cs^+$ ions. The spectra of the untreated LDPE (Fig. 7a,b) show hydrocarbon peaks (as $C_2H_2^-$, $C_3H_3^-$, and others) that are so much elevated that most of them appear saturated while the oxygenated peaks are present at lower intensity. In comparison, these oxygenated peaks from the treated LDPE (Fig. 7c,d) are higher than in the previous case. For example, the intensity of the $C_2HO^-$ ions at 10 nm increases from the untreated LDPE film (Fig. 7a) to the treated film (Fig. 7c), whereas the same intensity slightly decreases from Fig. 7b to Fig.7d for an in-depth of 40 nm. The increase in the intensity of oxygenated ions on a depth of 10 nm confirms the oxygen diffusion during the LDPE plasma treatment. The high intensity of $CHO_2^-$ (carboxylic ion) in both spectra of the treated film (Fig. 7c,d) is correlated with the increasing intensity of the carboxylic groups shown on the FTIR spectra (Fig. 4c). By comparing the depth profiling curves of the untreated and treated films, the oxygen diffusion depth could be assessed thanks to the oxygenated fragments.

For the sake of clarity and for the comparison with O/C XPS ratio, the overall oxygen content could be presented by the $OH^-/CH^-$ ratio shown in Fig. 8 versus the sputtering depth. Figure 8 shows the $OH^-/CH^-$ intensity values ratio as a function of the sputtering depth in the case of several LDPE films treated by an Ar post-discharge for (i) several treatment times (5, 10, 30, and 60 s), (ii) three plasma powers (60, 70, and 85 W), and (c) different oxygen flow rates (5, 15, and 25 mL.min$^{-1}$). Whatever the graph, the $OH^-/CH^-$ ratio increases rapidly in the first few nanometers of the profile and reaches a plateau for the films treated during a time shorter than 60 s (Fig. 8a) or with a power of 60W (Fig. 8b). In the other cases, the intensity increases rapidly to reach a maximum, immediately followed by a slow decrease until a plateau that is also reached by the other profiles.

| | Sampling depth analysis (nm) | | | | | | | | |
|---|---|---|---|---|---|---|---|---|---|
| $\theta$ | 18° | 25° | 35° | 45° | 55° | 65° | 75° | 85° | 90° |
| z (O 1s) | 2.3 | 3.2 | 4.3 | 5.4 | 6.3 | 6.9 | 7.4 | 7.6 | 7.6 |

*Table 2. X-ray photoelectron spectroscopy sampling depths as function of take off angle (measured from the surface)*

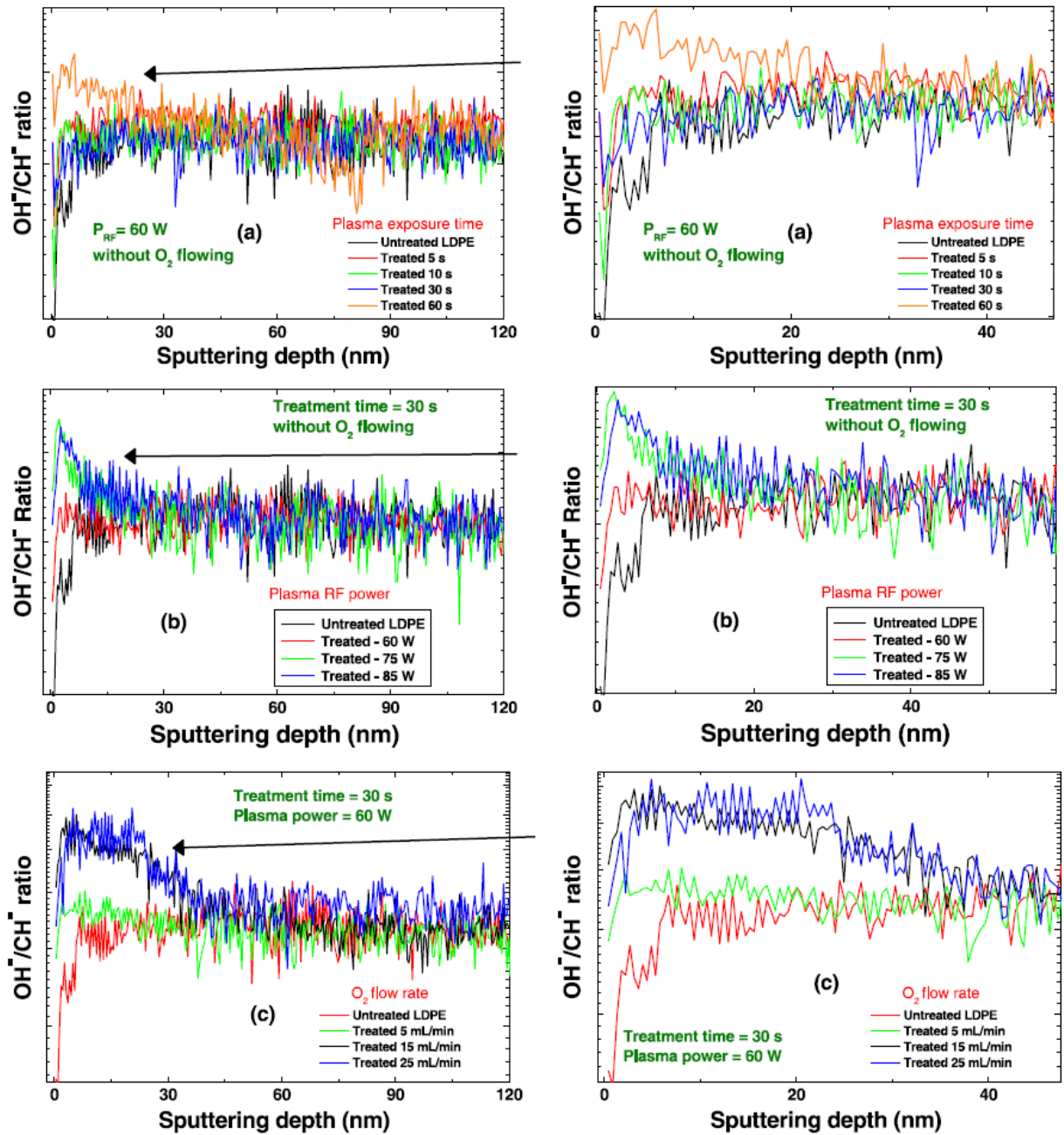

*Figure 8. ToF-SIMS OH⁻/CH⁻ ratio- depth profiles of untreated and treated LDPE layer using Cs⁺ sputtering beams at 750 eV as a function of (a) the treatment time; (b) the plasma power; and (c) the reactive oxygen flow.*

In Fig. 8a, the trend of the OH⁻/CH⁻ ratio in the case of the untreated film is quite similar with those of the treated films (5, 10, and 30 s). A different observation applies to Fig. 8b,c. Therefore, the OH⁻/CH⁻ ratio measured in this case corresponds only to the background signal already observed on the untreated LDPE films, thus indicating that a low oxygen concentration was already present in the untreated polymer. We notice that the oxygen diffuses slightly into the bulk if the films are treated in less than 60 s, or at a plasma power of 60 W, or with an oxygen flow rate as low as 5mL.min⁻¹. However, for LDPE films treated during 60 s (Fig. 8a) or for a power of 70 and 85W (Fig. 8b) or with oxygen flow rates higher than 5mL.min⁻¹ (Fig. 8c), the OH⁻/CH⁻ ratio for a sputtering depth comprised between 0 and 40nm remains quite elevated, followed by gradual decrease down to the level reached on the other profiles after 20–40nm of depth. As a consequence, if the conditions time >30 s, power >60 W, $O_2$ flow rate >5mL.min$^{-1}$ are respected, the oxygen diffusion becomes more significant into the bulk. The oxygen penetration depth, and thus the oxygen gas diffusion effect, can be estimated to about 40nm below the surface.

### 3.6. Study of the oxygen diffusion by AR-XPS

Because the difference in the signal intensities is not clear for the situations where the films were treated on short times (5, 10, or 30 s) or for a weak plasma power (60 W), the oxygen penetration depth was estimated on the first in-depth nanometers by AR-XPS. For plasma-treated polymer samples, the modification depth is typically of several hundreds of angstrom and was studied by surface-sensitive techniques such as AR-XPS.[30] AR-XPS was performed on the untreated and several treated LDPE surfaces for different times.

The estimated electron inelastic mean free path in the LDPE between 700 and 800 eV (kinetic energy of O1s) is comprised between 2.4 and 2.6nm[31] Table 2 represents the calculated analysis depths (nanometer)[32,33] by considering that the intensity on the top surface (with TOA equal to 18°) is the reference intensity, because the XPS analyzer limits are 15° and 90°.

The variation of the oxygen concentration in terms of the analysis depth (z) for the LDPE films treated at 60W and without adding oxygen gas is shown in Fig. 9. This figure clearly shows that the oxygen diffusion into the bulk of the polymer is dependent on the treatment time. The oxygen penetration depth is estimated to 8, 9, 10, and higher than 10 nm for the films treated during 5, 10, 30, and 60 s, respectively. We consider that there is no diffusion of oxygen atoms in the case of the untreated films, but the [O] remains almost low and constant whatever the TAO used.

These results relative to the oxygen penetration depth are complementary to those obtained using ToF-SIMS profiles (Fig. 8a) and clearly evidence that oxygen diffuses into LDPE even on short treatment times (5, 10, or 30 s).

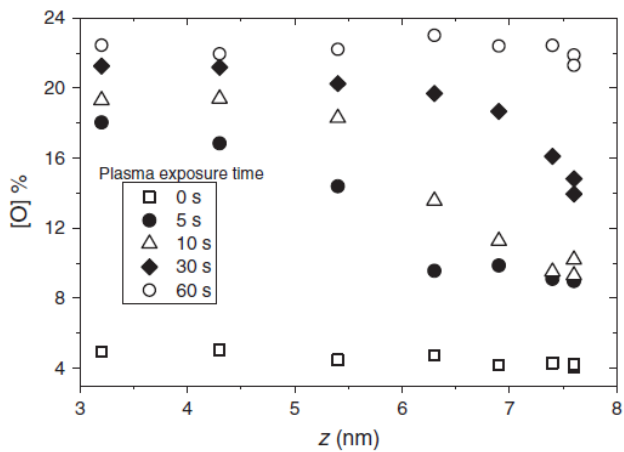

*Figure 9. Variation of the O concentration (%) as a function of the analysis depth, measured by AR-XPS. The LDPE films were treated at 60W in a pure Ar post-discharge.*

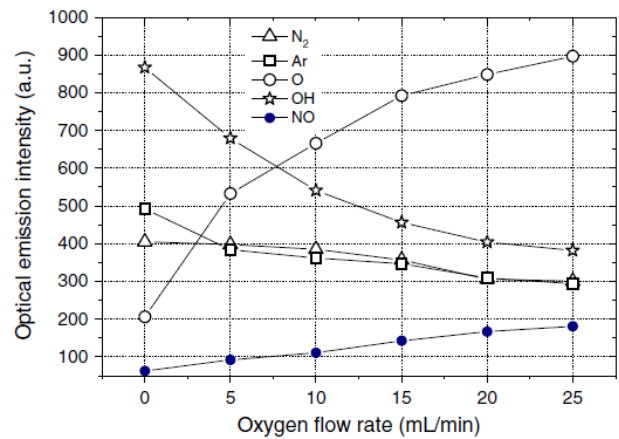

*Figure 10. Peak intensities of the observed species by OES versus the oxygen flow rate.*

### 3.7. Plasma diagnostics

Optical emission spectroscopy was used in order to determine the species that could be involved in the plasma surface modification depending on the oxygen flow rate. Figure 10 represents the intensities of the optical emission lines and bands of the following species as a function of the oxygen flow rate: O (777 nm), Ar (772 nm), $N_2$ (337 nm), OH (310 nm), and NO (279 nm). The emission band of $N_2$ is observed whatever the oxygen flow rate, hence suggesting a reaction between the flowing gas of the post-discharge and the surrounding air. As no N line was detected by OES, the production of the NO species could be explained according to the following reactions:

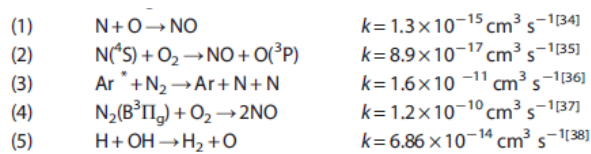

$$
\begin{aligned}
&(1) \quad N + O \rightarrow NO & k &= 1.3 \times 10^{-15} \text{ cm}^3 \text{ s}^{-1} [34] \\
&(2) \quad N(^4S) + O_2 \rightarrow NO + O(^3P) & k &= 8.9 \times 10^{-17} \text{ cm}^3 \text{ s}^{-1} [35] \\
&(3) \quad Ar^* + N_2 \rightarrow Ar + N + N & k &= 1.6 \times 10^{-11} \text{ cm}^3 \text{ s}^{-1} [36] \\
&(4) \quad N_2(B^3\Pi_g) + O_2 \rightarrow 2NO & k &= 1.2 \times 10^{-10} \text{ cm}^3 \text{ s}^{-1} [37] \\
&(5) \quad H + OH \rightarrow H_2 + O & k &= 6.86 \times 10^{-14} \text{ cm}^3 \text{ s}^{-1} [38]
\end{aligned}
$$

In the post-discharge, $N_2$ could be dissociated by colliding with Ar* species, as suggested in reaction (3) and evidenced in a previous work.[39] This reaction would result into the production of nitrogen atoms that could either recombine with O atoms according to reaction (1) or – in a less extent– collide with $O_2$ molecules according to reaction (2) to form the NO species. Another reaction responsible for the production of the NO species would be the recombination of $N_2(B^3\Pi_g)$ and $O_2$ into NO molecules, as stated by reaction (4). In Fig. 10, the increase in O and NO intensities as a function of the oxygen flow rate is consistent with these assumptions.

The decrease in the OH band is balanced by the increase in the O singlet line suggesting a consumption of OH by hydrogen to produce atomic oxygen (Reaction 5), even if the species (H or $H_2$) involving in this reaction which is taking place in the discharge were not detectable by OES. Figure 10 shows also the intensity of Ar singlet (14.7 eV of upper energy level) line slightly decreasing with the oxygen flow rate, because the molar fraction of the reactive gas is much smaller than the argon.

## 4. Discussion

The treatment of LDPE with an argon plasma torch induces a decrease in the WCA (40–50°) compared with the untreated film. The fact that the hydrophilicity of a treated film is better in the case of an Ar-$O_2$ post-discharge (plateau = 40°) than for a pure Ar post-discharge (plateau = 50°), could be attributed to a slight etching of the surface topography in the first case. During the treatment, the free radical sites produced on the polymer surface can react with the surrounding polymer molecules and the plasma-phase species. Morajev et al.[40] reported that the main species present in the afterglow of argon and oxygen plasma torch were the ground-state O atoms, ozone, metastable molecules ($O_2(^1\Delta_g)$, $O_2(^1\Sigma_g^+)$), oxygen radicals ($O(^1D)$), ions ($OH^-$), free electrons, and UV radiations. The authors[40] asserted the production of $1.2\pm0.4*10^{17}$ cm$^{-3}$ of ground-state oxygen atoms. Figure 10 confirms the presence of O atoms in this study and their increase with the oxygen flow rate. These species especially the oxygen atoms can react with the polymer chains to produce radicals on the polymer surface.

The collisions between the excited argon metastable species and the $O_2$ molecules result in an energy transfer to the reactive gas molecules. The chemical reactions between a polymer surface and an argon post-discharge can involve the abstraction of hydrogen atoms and the subsequent formation of carbon radicals.[41] An oxygen atom (10.74 eV of upper energy level) can easily abstract an H atom and therefore break a C–C bond (bond dissociation energies of C–H and C–C: 4.25 and 3.8 eV, respectively).

Even if the emission of the NO species remains low whatever the oxygen flow rate, it cannot be considered as negligible because some nitrogen functions ([N1s] ~ 1.5% for TOA = 15° only) were grafted onto the LDPE surface, as explained in the literature.[42] This proves that the nitrogen grafting concerns the first layer of the film and probably does not diffuse deeper as the oxygen. In the case of a PTFE treatment with a plasma torch supplied in helium, Dufour[43] et al. found no interaction between the post-discharge and the surrounding air provided a gap lower than 2 mm.

The increase in the hydrophilicity and in the O/C ratio of the LDPE films plotted in Figs. 2 and 3 as a function of the oxygen flow rate, the treatment time, and the plasma power may result from the increase in the atomic oxygen concentration in the post-discharge. Once the surface reached the plateau, this hydrophilicity is no more enhanced by increasing the treatment time, the RF power, or the $O_2$ flow rate. A similar effect was observed on PET, even if in that case, the initial situation was characterized by an enrichment of oxygen in the topmost surface layers.[44] In our work, the surface of the films starts to be saturated in oxygenated groups either for a long treatment time or with a high power or with an elevated oxygen flow rate. This surface saturation, evidenced by a WCA and O/C ratio results, is gradually followed by oxygen diffusion as a function of the treatment time. The AR-XPS and ToF-SIMS measurements shown in Figs. 8 and 9 are complementary and assess a diffusion of oxygen on a depth comprised between 2 and 40 nm, depending on the plasma conditions.

The following steps, from the first microseconds to the end of the treatment, can be summarized as presented in Fig. 11: the activation of the surface by oxygen atoms, the surface saturation, and the oxygen diffusion. On the one hand, as the hydrophilicity of the surface reaches a plateau (50° or 40° in an Ar or in an Ar-$O_2$ post-discharge), the top surface (first layer of LDPE) could be saturated after 30 s of treatment by oxygen groups until the oxygen atoms diffuse into the bulk to reach the second monolayer (a). On the other hand, there is no evidence, by technical surface analysis, that oxygen atoms saturate the first monolayer of the LDPE surface. This suggests also that oxygen atoms could diffuse toward the monolayers from the first microseconds of the plasma treatment (b). This behavior needs more critical investigation because a smooth etching process could occur during the plasma treatment. The analysis of the first monolayer of a treated LDPE and the smooth etching process is intended in the framework of a future study.

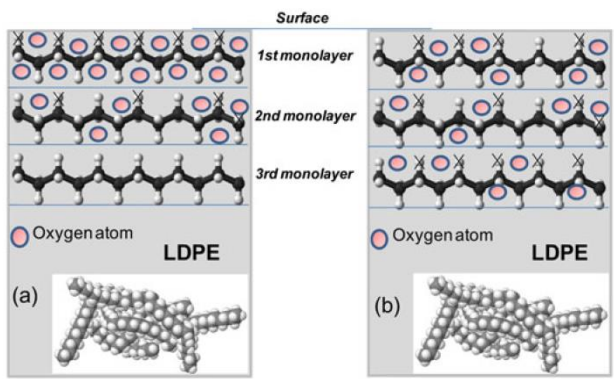

Figure 11. Diagram illustrating the LDPE plasma treatment with (a) saturation and (b) non-saturation of the first monolayer.

The variation of the oxygen penetration depth in terms of the plasma parameters is represented in Fig. 12. For a film treated with a pure Ar post-discharge (plasma power = 60W) during 60 s, this depth could be evaluated to about 27nm (Fig. 12a), whereas with an Ar-$O_2$ post-discharge (oxygen flow rate = 25mL.min$^{-1}$) during 30 s, this depth could be evaluated to 40nm (Fig. 12c). The modification depth can be estimated between 10 and 40nm depending on the plasma parameters and the nature of the LDPE polymer. For instance, a spin-coated HDPE deposited on silicon substrate, treated 2000 s at low pressure in Ar plasma,[45] presents a modification depth turning around 50nm below the surface. Because LDPE presents more ramifications in its structure than HDPE, the chains are less cross-linked, hence, a lower intensity of the Van der Waals intermolecular forces. This nature and the fact that the LDPE is more amorphous (63%) permits the oxygen species to penetrate deeper in the bulk of the LDPE after tens of seconds of the plasma treatment. The ToF-SIMS results clearly showed that, compared with the untreated PE films, the plasma treatment induced an increase in the OH$^-$/CH$^-$ ratio, which is very consistent with the XPS and FTIR results.

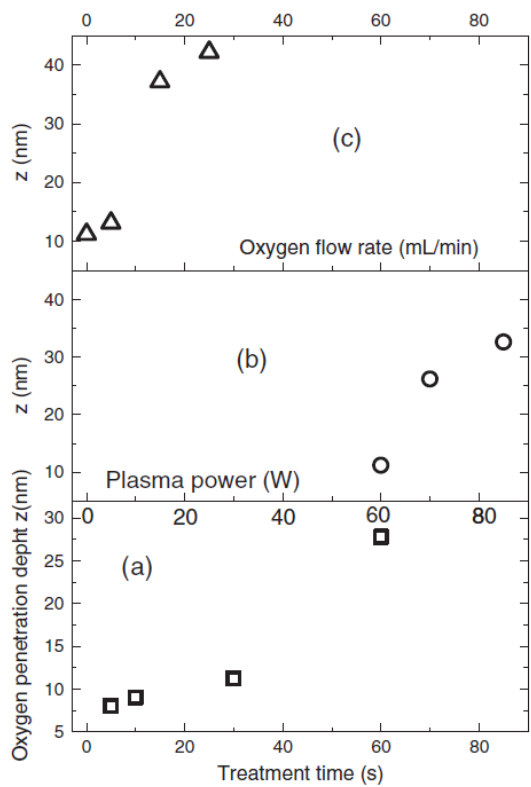

Figure 12. Oxygen penetration depth z (nanometer) in terms of (a) the treatment time (s); (b) the plasma power (W) without oxygen gas for a treatment time of 30 s; and (c) the oxygen flow rate for a treatment time of 30 s and a power of 60 W.

## 5. Conclusion

This article describes the oxygen diffusion into LDPE films induced either by an Ar or an Ar-$O_2$ post-discharge. The wettability of the surface film reaches a plateau (around 50°) when the treatment time is equal to 120 s and a plateau (around 40°) when the oxygen flow rate is higher than 5mL.min$^{-1}$. This plateau, suggesting a saturation of the oxygen content on the surface, is followed by an oxygen diffusion toward the bulk. The oxygen diffusion was also confirmed by FTIR analysis. On the basis of the ToF-SIMS profiles and the AR-XPS measurements, the oxygen depth penetration was estimated from 1 to 40 nm, depending on the plasma parameters. The atomic oxygen and NO species detected by OES are the main reactive species that could be responsible for the surface activation and the oxygen diffusion.

## 6. Acknowledgements

We gratefully acknowledge the support of the Everwall Project financed by the FEDER – Région Wallonne and Europe. F. Reniers also thank the I.A.P. 'Physical chemistry of plasma surface interactions-PSI' was funded by the Belgian Federal Government.

## 7. References

[1] D. Hegeman, H. Brunner, C. Oehr, Nucl. Instr. Meth. Phys. Res. 2003, 208, 281.

[2] R. Seebock, H. Esrom, M. Charbonnier, M. Romand, U. Kogelschatz, Surf. Coat. Technol. 2001, 142, 455.

[3] L. K. Pochner, S. Beil, H. Horn,M. Blomer, Surf. Coat. Technol. 1997, 97, 372.

[4] H. Kaczmarek, J. Kowalonek, A. Szalla, A. Sionkowska, Surf. Sci. 2002, 507, 883.

[5] U. Kogelschatz, IEEE Trans. Plasma Sci. 2002, 30, 1400.

[6] C. Z. Liu, N. Y. Cui, N. M. D. Brown, B. J. Meenan, Surf. Coat. Technol. 2004, 185, 311.

[7] M. Morajev, R. F. Hicks, Chem. Vap. Depos. 2005, 11, 469.

[8] W. Chen, C. Jie-rong, L. Ru, Appl. Surf. Sci. 2008, 254, 2882.

[9] L. Ru, C. Jie-rong, Appl. Surf. Sci. 2006, 252, 5076.

[10] N. Dumitrascu, G. Borcia, N. Apetroaei and G. Popa, Plasma Sources Sci. Technol. 2002, 11, 127.

[11] C.-S. Ren, K. Wang, Q. Y. Nie, D. Z. Wang, S. H. Guo, Appl. Surface Sci. 2008, 255, 3421.

[12] G. Borcia, A. Chiper, I. Rusu, Plasma Sources Sci. Technol. 2006, 15, 849.

[13] D. Pappas, A. A. Bujanda, J. A. Orlicki, R. E. Jensen, Surf. Coat. Technol. 2008, 203, 830.

[14] M. A. Gilliam, Q. S. Yu, J. Appl. Polymer Sci. 2006, 99, 2528.

[15] S.Wu, Polymer Interface and Adherence,Marcel Dekker, NewYork, 1982.

[16] V. Svorcik, K. Kolarova, P. Slepicka, A. Mackova, M. Novotna, V. Hnatowicz, Polym. Degrad. Stabil. 2006, 91, 1219.

[17] B. Nisol, C. Poleunis, P. Bertrand, F. Reniers, Plasma Process. Polym. 2010, 7, 715.

[18] C. D. Wagner, W. M. Riggs, L. E. Davis, J. F. Moulder, G. E. Muilenberg, Handbook of X-ray Photoelectron Spectroscopy (1st edn), Perkin Elmer Corporation, Physical Electronics Division, Germany, 1979.

[19] U. Schulz, P. Munzert, N. Kaiser, Surf. Coat. Technol. 2001, 142, 507.

[20] E. Gonzalez II, R. F. Hicks, Langmuir 2010, 26, 3710.

[21] G. Borcia, C. A. Anderson, N. M. D. Brown, Appl. Surf. Sci. 2004, 221,203.

[22] N. Encinas, B. Díaz-Benitoa, J. Abenojara, M. A. Martínez, Surf. Coat. Technol. 2010, 205, 396.

[23] M. Ataeefard, S. Moradian, M. Mirabedini, M. Ebrahimi, S. Asiaban, Prog. Organic Coat. 2009, 64, 482.

[24] M. R. Sanchis, V. Blanesa, M. Blanesa, D. Garciab and R. Balart, Europ. Polymer J. 2006, 42, 1558.

[25] N. Vandencasteele, F. Reniers, Surf. Interface Anal. 2004, 36, 1027.

[26] M. Lehocky, H. Drnovska, B. Lapcıkova, A. M. Barros-Timmons,

T. Trindade, M. Zembala, L. Lapcik, J. Col. Surf. A: Physicochem. Eng. Aspects 2003, 222, 125.

[27] A. G. Shard, F. M. Green, P. J. Brewer, M. P. Seah, I. S. Gilmore, J. Phys. Chem. B 2008, 112, 2596.

[28] L. Houssiau, N. Mine, Surf. Interface Anal. 2010, 42, 1402.

[29] N. Mine, B. Douhard, J. Brison, L. Houssiau, Rapid Commun. Mass Spectrom. 2007, 21, 2680.

[30] C. M. Chan, Polymer Surface Modification and Characterization, Hanser, New York, 1994.

[31] S. Tanuma, C. J.Powell, D. R. Penn, Surf. Interface Anal. 1994, 21, 165.

[32] C. M. Chan, Polymer Surface Modification and Characterization, Hanser Publishers, New York, 1993.

[33] C. Jama, J. D. Quensierre, L. Gengembre, V. Moineau, J. Grimblot, O. Dessaux, P. Goudmand, Surf. Interface Anal. 1999, 27, 653.

[34] E. C. Zipf, Can. J. Chem. 1969, 47(10), l863.

[35] M. Morajev, X. Yang, M. Barankin, J. Penelon, S. E. Babayan, R. F. Hicks, Plasma Sources Sci. Technol. 2006, 15, 204.

[36] A. J. Barnett, G. Marston, R. P. Wayne, J. Chem. Soc. Faraday Trans. 1987, 83, 1453.

[37] L. G. Piper, J. Chem. Phys. 1992, 97, 270.

[38] M. J. Kirkpatrick, B. Dodet, E. Odic, Intern. J. Plasma Envi. Sci. Technol. 2007, 1, 96.

[39] C. Y. Duluard, T. Dufour, J. Hubert, and F. Reniers, J. Appl. Phys. 2013, 113, 93303.

[40] M. Moravej, X. Yang,; R. F. Hicks, J. Penelon, S. E. Babayan, J. Appl. Phys. 2006, 99, 93305.

[41] F. Clouet, M. K. Shi, J. Appl. Polym. Sci. 1992, 46, 1955.

[42] A. J. Wagner, D. H. Fairbrother, F. Reniers, Plasmas Polym 2003, 8, 119.

[43] T. Dufour, J. Hubert, N. Vandencasteele, F. Reniers, Plasma Sources Sci. Technol. 2012, 21, 45013.

[44] C. Lopez-Santos, F. Yubero, J. Cotrino, A. R. Gonzalez-Elipe, ACS Appl. Mater. Interf. 2010, 2, 980.

[45] S. K. Øiseth, A. Krozer, B. Kasemo, J. Lausmaa, Appl. Surf. Sci. 2002, 202, 92.